\documentclass[letter,scriptaddress,twocolumn,prl,showkeys,showpacs,superscriptaddress]{revtex4-2}

	\usepackage{amsmath}%,amssymb} 
	\usepackage{makeidx}
	\usepackage{amsfonts}
	\usepackage{mathrsfs}
	\usepackage{bbm}
	\usepackage{graphicx, xcolor}  % needed for figures
	\usepackage{dcolumn}   % needed for some tables
	\usepackage{bm}        % for math
	\usepackage{amssymb}   % for math
	\usepackage[ansinew]{inputenc} 
	\usepackage[usenames,dvipsnames]{pstricks}
	\usepackage{subfigure}
	\usepackage{epstopdf}
	\usepackage{pst-grad} % For gradients
	\usepackage{pst-plot} % For axes
	\usepackage[colorlinks,hyperindex]{hyperref}
	\hypersetup
	{
		colorlinks,%
		citecolor=black,%
		linkcolor=black,%
		urlcolor=black,%
	}

\newcommand{\be}{\begin{equation}}
\newcommand{\ee}{\end{equation}}
\newcommand{\bse}{\begin{subequations}}
\newcommand{\ese}{\end{subequations}}
\newcommand{\ba}{\begin{aligned}}
\newcommand{\ea}{\end{aligned}}
\newcommand{\bsp}{\begin{split}}
\newcommand{\esp}{\end{split}}
\newcommand{\bme}{\begin{multline}}
\newcommand{\eme}{\end{multline}}

\begin{document}

\title{Molecular tug of war reveals adaptive potential of an immune cell repertoire}
%\title{Physical constraints reveal adaptive potential of immune learning}
\author{Hongda Jiang}
\author{Shenshen Wang}
\email{shenshen@physics.ucla.edu}
\affiliation{Department of Physics and Astronomy, University of California, Los Angeles, Los Angeles, CA 90095}

%\date{\today}

%\maketitle

\begin{abstract}
The adaptive immune system constantly remodels its lymphocyte repertoire for better protection against future pathogens. Its ability to improve antigen recognition on the fly relies on somatic mutation and selective expansion of B lymphocytes expressing high-affinity antigen receptors. However, this Darwinian process inside an individual appears ineffective, hitting a modest ceiling of antigen-binding affinity. Experiment began to reveal that evolving B cells physically extract antigens from presenting cells and that the extraction level dictates clonal expansion; this challenges the prevailing assumption that the equilibrium constant of receptor-antigen binding determines selective advantage. We present a theoretical framework to explore whether, and how, such tug-of-war antigen extraction impacts the quality and diversity of an evolved B cell repertoire. We find that the apparent ineffectiveness of clonal selection can be a direct consequence of the non-equilibrium nature of antigen recognition. Our theory predicts that the physical strength of antigen tethering under tugging forces sets the affinity ceiling. Meanwhile, the model showed that, intriguingly, cells can use force variability to diversify binding phenotype without compromising fitness, thus remaining plastic under resource constraint. These results suggest that active probing of receptor quality via a molecular tug of war during antigen recognition limit the potency of response to the current antigen, but confer adaptive benefit for protection against future variants. Importantly, a saddle point in the fitness landscape of B cell-phenotype evolution emerges from the tug-of-war setting, which rationalizes multiple key phenomenology and puts forward a role of active physical dynamics in immune adaptation.
\end{abstract}

\maketitle 

\section{Introduction}
The adaptive immune system protects living organisms against a vast and changing variety of microscopic invaders. Adaptive immunity relies on dynamic reorganization of populations of B and T lymphocytes that express unique antigen receptors on their surface to recognize and remember encountered pathogens. The ability of an immune cell repertoire to learn from past encounters and improve antigen recognition on the fly -- one of the hallmarks of natural immunity -- is crucially dependent on somatic hypermutation~\cite{di2007} and selective expansion of B cells expressing high-affinity antigen receptors. Upon recognition of a pathogen, the process of affinity maturation~\cite{victora2012} -- a form of rapid Darwinian evolution within an individual -- allows a diverse pool of naive (antigen-inexperienced) B cells to iteratively enhance the binding affinity of their receptors to the infecting antigen, generating a pool of memory cells with varied lifetimes and long-lived plasma cells that secrete antibodies (membrane-detached B cell receptors).  

Since specific molecular interactions mediate immune recognition, previous models of affinity maturation often assume that \textit{equilibrium} binding affinity between B cell receptor (BCR) and antigen (Ag) determines the reproductive success of a B cell clone~\cite{zhang2010, wang2015, childs2015, shaffer2016, wang2017}. However, \textit{in vivo} affinity maturation appears rather ineffective: First, it retains B cell clones with a wide variety of binding affinities for the current antigen~\cite{eisen1964} and produces memory B cells of low specificity and high diversity~\cite{kuraoka2016, nakagawa2021}. More importantly, the B cell response exhibits a modest ceiling of antigen-binding affinity, characterized by an equilibrium dissociation constant $K_\mathrm{d}>0.1\, \mathrm{nM}$~\cite{batista1998, foote1995}. In contrast, \textit{in vitro} evolved antibody mutants (via directed evolution) can achieve a monovalent binding affinity with $K_\mathrm{d}\sim 50\, \mathrm{fM}$~\cite{boder2000}, more than three orders of magnitude above the \textit{in vivo} affinity ceiling. The existence of these mutants demonstrates that B cell receptors are not intrinsically responsible for an affinity ceiling, pointing toward \textit{in vivo} constraints not present in \textit{in vitro} settings.   

\textit{In vitro} measurements of binding affinity take place after equilibration is reached, and binding-curve fitting is based on the Langmuir isotherm~\cite{soltermann2021}. Yet, \textit{ex vivo} experiments have started to reveal that cells do their affinity measurements differently: Evolving B cells exert mechanical pulling forces generated by the actin cytoskeleton to physically extract antigens from the antigen presenting cells (APCs), prior to internalization through endocytosis~\cite{nowosad2016, spillane2017}. In addition, antigens are attached to the APCs via protein tethers of various types: some are long and flexible and others short and stiff; even newly produced antibodies can serve as tethers by forming immune complexes with antigens~\cite{zhang2016, toellner2018}. Thus, B cells evolving \textit{in vivo} actively probe their receptor quality during antigen recognition through a tug-of-war extraction process; both active force usage and tether properties (factors extrinsic to BCR) may influence the extraction outcome. Furthermore, intravital imaging has indicated that, during an immune response, the amount of antigen a B cell acquires and subsequently presents to T helper cells determines its number of offspring~\cite{gitlin2014}. In this way, B cells translate antigen-binding affinity into clonal reproductive fitness, through the efficiency of antigen extraction. Connecting this intriguing set of observations, we propose that the apparent ``ineffectiveness" of selection is not an artifact due to inevitable randomness, but rather, can be a direct, functional consequence of antigen recognition being \textit{out of equilibrium}. This challenges prevailing model assumptions based on equilibrium binding and raises the need for assessing what microscopic aspects of non-equilibrium recognition may yield a macroscopic impact on evolution of B cell responses. 

Here we present a theoretical framework to explore ways in which non-equilibrium antigen recognition may influence the quality and diversity of a B cell repertoire. We developed a simple stochastic model of tug-of-war antigen extraction that maps equilibrium binding free energies to extraction probability. This mapping provides the bridge between binding affinity and reproductive fitness which then forms the central piece (a force- and tether-modulated affinity-dependent proliferation rate) of an iterative program mimicking affinity maturation \textit{in silico}. We find that tug-of-war antigen extraction indeed limits evolvable antibody affinity; theory predicts an affinity ceiling in terms of effective tether strength under tugging forces. 
%Intuitively, once the receptor-antigen bond exceeds the tether strength, B cells of even higher affinities have no more selective advantage, since bond rupture almost always disrupts the tethering complex, resulting in antigen extraction. 
Then, why do cells expend energy to limit their own evolution? Our theory suggests that tug-of-war extraction may in fact represent an adaptive strategy: heritable heterogeneity of pulling force magnitude, combined with evolvable receptor flexibility, is found to be able to generate a wide variety of binding affinities with similar clonal fitness. Such phenotypic plasticity may allow a balance between depth and breadth of protection against evolving pathogens with a finite repertoire. This result provides a testable alternative to the empirical speculation that \textit{in vivo} affinity should be limited to ensure response specificity and avoid auto-immunity. Our theory further shows that, interestingly, the extraction system generically exhibits a saddle point in the fitness landscape of B cell-phenotype evolution, a feature that naturally recapitulates multiple key experiments otherwise hard to reconcile, such as retention of low-affinity clones~\cite{eisen1964, kuraoka2016, nakagawa2021} and widely disparate rates of diversity loss among B cell populations~\cite{tas2016}.       
Lastly, our theory generates independent predictions that would allow experiment to verify or falsify the proposal.
%including a non-monotonic dependence of evolved phenotypic diversity on initial force heterogeneity

Our work is complementary to most of the modeling efforts in the field. 
Statistical models of immune systems have explored how resource costs and evolutionary constraints influence immune responses to diverse and evolving pathogens~\cite{mayer2015, mayer2019, chardes2022, marchi2021, cobey2017, sachdeva2020, jones2021, sheng2021, bradde2020, luo2015, amitai2018}. From a complementary perspective, we will demonstrate that active forces and physical constraints can be harnessed to promote the emergence of phenotypic plasticity.
In a broader context, theoretical work on biological adaptation highlight the impact of environmental statistics, correlation structure and variational timescales on adaptation strategies~\cite{kussell2005, rivoire2011, xue2019, tikhonov2020, mayer2016, kashtan2007, parter2008, mustonen2010, wang2019, murugan2021}. Our framework, instead, aims to characterize the role of physical dynamics of cells during signal acquisition in shaping the rapid evolution of adaptive responses. 

\section{Model}

\begin{figure*}
\includegraphics[width=0.96\textwidth]{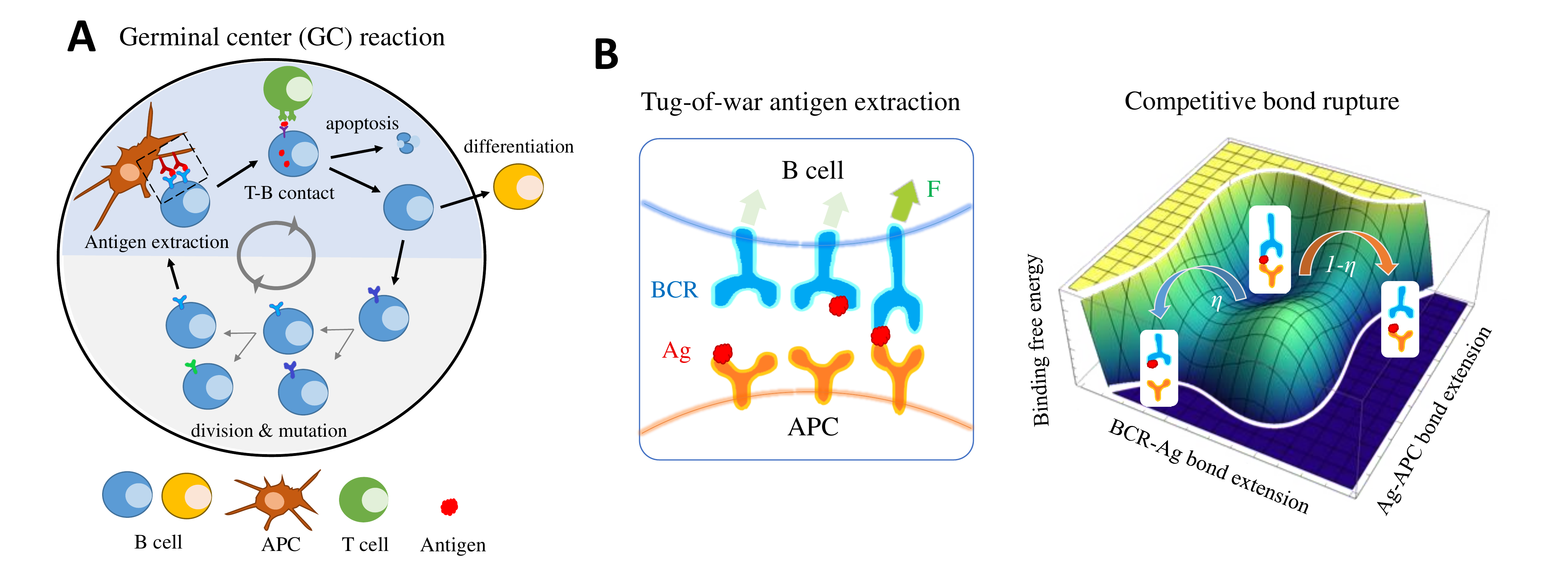}
\caption{An overview of the model framework: evolutionary learning of B cell repertoire through affinity maturation crucially depends on physical extraction of antigen via a molecular tug of war.
%tug-of-war antigen extraction maps receptor properties to clonal fitness.
(A) Affinity maturation takes place in germinal centers (GCs) -- structured microenvironments where cellular players meet and interact. Iterative evolutionary learning proceeds via GC cycles: somatic hypermutations cause changes in BCR properties (affinity and flexibility) during proliferation which alter the efficiency of B cells to acquire antigen from the surface of antigen presenting cells and subsequently present to helper T cells. This efficiency, in turn, determines the selective advantage of competing cells. Positively selected B cells (avoiding apoptosis) either differentiate into memory or plasma cells and exit the GC, or join the next cycle of reactions.   
(B) The central element of this evolutionary program is the relationship between receptor-antigen binding characteristics and reproductive fitness of a B cell, where the key is nonequilibrium extraction dynamics via competitive bond rupture under pulling force, i.e. stochastic escape from the bound state over one of the activation barriers (right panel). Here, an BCR-Ag-APC complex -- consisting of two binding interfaces -- is a much simplified model of the protein chain linking a B cell to the APC and yet captures the essence of a tug-of-war configuration.
The Ag-APC bond combines Ag-tether and tether-APC associations. Tether (orange) can be fixed or renewable.
APC: antigen presenting cell, Ag: antigen, BCR: B cell receptor.}
\label{model}
\end{figure*}

Current knowledge of somatic evolution paints a highly dynamic picture of repertoire remodeling:
Upon activation due to antigen recognition, B cells proliferate to form populations of thousands of cells in germinal centers (GCs), where they mutate during replication and compete fiercely for survival (Fig.~\ref{model}A). These compartmentalized microenvironments allow frequent encounter of B cells and helper T cells in the midst of space-spanning networks of follicular dendritic cells; these GC-resident antigen presenting cells display antigens on their surface, mediate receptor-antigen interaction, and support antigen extraction through cell-cell contact.

%We seek to understand whether immune cells gain any functional advantage by actively interacting with the physical environment, and if so, under what conditions. Of particular interest are the ways in which active force usage during antigen acquisition influences the quality and diversity of the evolved receptor repertoire. To probe possible macroscopic consequences of microscopic processes, we attempt to link the free energy landscape of receptor-antigen binding to reproductive success of cells, through physical dynamics of antigen extraction (section A). This mapping, in turn, forms the central piece (force-modulated affinity-dependent proliferation rate) of an iterative program of B cell competitive evolution mimicking affinity maturation \textit{in silico} (section B). We will show that this physical theory of antigen extraction not only accounts for natural limits of B cell selection, but also reveals evolutionary goals and design principles distinct from expectations based on equilibrium binding alone. 

\subsection{A. Stochastic antigen extraction via a molecular tug of war}

Antigen extraction occurs through competitive rupture of the BCR-Ag bond and the Ag-APC bond under pulling force in a tug-of-war configuration (Fig.~\ref{model}B); survival of the BCR-Ag bond beyond the lifetime of the Ag-APC bond leads to a successful extraction event. Here the Ag-APC bond is a coarse-grained description of the potentially complex tethering interaction (including the antigen-tether attachment and the associated APC membrane).
Extraction is stochastic in nature, because bond rupture occurs via activated dynamics, i.e., thermally assisted escape from the bound state over an activation barrier. Extending the Kramers theory~\cite{hanggi1990} to a chain of adhesive bonds, we formulate the extraction dynamics with coupled Langevin equations describing the motion of antigen and BCR, respectively, in the overdamped limit (see SI text)    
\begin{eqnarray}
\label{EOM}
&\gamma_a \dot{x}_a = -U_a'(x_a) + U_b'(x_b) +\xi_a,\\ \nonumber
&\gamma_b (\dot{x}_a + \dot{x}_b) = -U_b'(x_b) +F + \xi_b.
\end{eqnarray}
Here $U_a(x_a)$ and $U_b(x_b)$ are profiles of intrinsic binding free energy for the Ag-APC and BCR-Ag bonds along the pulling direction, with bond extensions $x_a$ and $x_b$ being the reaction coordinates. Our results hold qualitatively for any smooth single-well profile, characterized by an activation-barrier height $\Delta G^\ddagger$ and a bond length (i.e. minimum-to-barrier distance) $x^\ddagger$. 
Pulling force $F$ and tether properties $\{\Delta G_a^\ddagger,\, x_a^\ddagger\}$ can be time dependent, leading to a dynamic fitness landscape in the space of BCR traits $\{\Delta G_b^\ddagger,\, x_b^\ddagger\}$ (Results section C). 
Fandom forces $\xi$ and frictional forces $-\gamma\dot{x}$ both arise from collisions with particles in the ambient fluid and are related through the fluctuation-dissipation theorem, $\langle \xi_i(t)\xi_j(t')\rangle=2k_B T\gamma_i\delta_{ij}\delta(t-t')$, with $i, j=a, b$; $\langle\xi_i\rangle=0$. Damping coefficients $\gamma_i$'s set relaxation timescales.
Thus, these equations of motion depict antigen extraction under frictional, elastic, pulling and random forces. 

To quantify the intuition of competitive bond rupture, we calculate the extraction probability $\eta$ as the chance by which the BCR-Ag bond persists longer than the Ag-APC bond. In the limit of high activation barriers, a simple factorized form results:
\begin{equation}\label{eta}
\eta=\int_0^{\infty}\mathrm{d}t p_a(t)S_b(t), 
\end{equation}
where $S_b(t)=\int_t^{\infty}\mathrm{d}t' p_b(t')$ is the survival probability of the BCR-Ag bond until at least time $t$ when the Ag-APC bond breaks. $p_a(t)$ and $p_b(t)$ are distributions of bond lifetime governed by Eq.~\ref{EOM} (i.e. first passage time to exceeding rupture lengths $x_a^\ddagger$ and $x_b^\ddagger$ that act as absorbing boundaries).
Under modest constant pulling forces, bond lifetimes are nearly exponentially distributed and the extraction probability simply is $\eta = 1/(1+\tau_a/\tau_b)$; here $\tau_a$ and $\tau_b$ are mean first passage times conditioned on exiting through the Ag-APC and the BCR-Ag rupture boundary, respectively, treating the other boundary as reflective. This relationship suggests that the extraction system implement a nonequilibrium ratio test of dynamic bond strengths; by counting successful events out of many extraction attempts, cells can measure the ratio of mean lifetimes between tugging and tethering bonds. 
%Extraction probability then provides the bridge between molecular interaction and clonal selection (Eq.~\ref{fitness}).

In this work, we consider a mean-field picture of antigen extraction, assuming independent BCR-Ag-APC complexes subject to equal pulling stress. 
%Specifically, we assume that pulling stress spreads evenly across individual BCR-Ag-APC binding complexes and treat extraction events as independent. 
Admittedly, there is a gap between individual extraction events and the overall extraction level of a cell; the link should be built by accounting for dynamic organization of the B cell-APC contact pattern (immunological synapse) through which antigens are extracted. We will outline in Discussion how this might be implemented, but proceed with the mean-field model hereafter for two reasons. First, the assumption is compatible with the observation that traction force applied to a receptor cluster scales with its size~\cite{tolar2014, wang2018}. More importantly, as we will show, that the mean-field picture can already rationalize multiple key phenomenology supports the molecular tug of war (not yet modulated by cellular processes) as a plausible microscopic mechanism essential for understanding repertoire responses.

\subsection{B. Germinal-center reaction implements evolutionary learning of a B cell repertoire} 

Affinity maturation manifests itself as an \textit{in vivo} evolutionary optimization of molecular recognition, because the efficiency of antigen acquisition -- a nonlinear function of evolvable receptor traits -- was found to dictate the proliferation rate of a B cell~\cite{gitlin2014}. It thus follows that force can shape the selection pressure by modulating the extraction likelihood $\eta$. To characterize in what manner force usage influences evolutionary outcomes, e.g. receptor potency and diversity, we build a birth-death-mutation model of GC reaction (Fig.~1A) and simulate the reaction cycles using stochastic agent-based algorithms (SI text). 

Motivated by the observation of affinity-based proliferation~\cite{nakagawa2021}, we model the growth rate as $r(\eta)=\lambda(\eta)\left(1-N/N_c\right)$, where the carrying capacity $N_c$ accounts for space and resource constraints on affinity maturation, and the proliferation rate $\lambda(\eta)$ depends on binding affinity through extraction probability
\begin{equation}\label{fitness}
\lambda(\eta) = \lambda_0 \frac{n_{\rm Ag}(\eta) }{n_0 + n_{\rm Ag}(\eta)}.
\end{equation}
This sigmoidal dependence on the amount of extracted antigen, $n_{\rm Ag}$, is chosen to approximate a global nonlinearity that summarizes intracellular processing of extracted antigen and subsequent competition for T cell help~\cite{crotty2015}. Assuming independent extraction events (mean-field assumption), we draw $n_{\rm Ag}$ from a binomial distribution $n_{\rm Ag}\sim B(C_\mathrm{Ag} A, \eta)$, where $C_\mathrm{Ag}$ is the surface concentration of BCR-Ag-APC complexes and $A$ the contact area between a B cell and the APC, both being treated as affinity independent. 
The extraction level $n_0$ leading to half-maximum replication rate sets a threshold for B cell survival. A death rate is assumed to be constrained and fixed, uniform among cells. 

Intrinsic parameters of a BCR -- binding affinity $\Delta G_b^\ddagger$ and bond length $x_b^\ddagger$ -- characterize its binding free energy surface for a given antigen in the absence of force. Both traits can be modified through somatic mutations that introduce point changes to BCR-encoding gene segments; changes in the complementarity-determining regions directly influence antigen binding ($\Delta G_b^\ddagger$), whereas changes to the framework regions mainly alter BCR flexibility ($x_b^\ddagger$). 
We assume that upon mutation, each parameter picks up an increment according to a Gaussian distribution with a typical jump size $\sigma$:
\begin{eqnarray}\label{evo}
\Delta G^\ddagger_{b,\,\, t+1}  &= \Delta G^\ddagger_{b, \,\, t} + \sigma_{G_b}\eta_{t+1},\\ \nonumber
x^\ddagger_{b,\,\, t+1}  &= x^\ddagger_{b, \,\, t} + \sigma_{x_b}\xi_{t+1},
\end{eqnarray}
where $\langle \eta_t \rangle = \langle \xi_t \rangle=0$ and $\langle \eta_t \eta_{t'} \rangle = \langle \xi_t \xi_{t'} \rangle = \delta_{t,\,t'}$. This form of mutation-induced changes allows both traits to continually evolve. Thus, if increases in affinity were to slow down within the model, it must be due to factors other than a lack of beneficial mutations.

An evolutionary learning (GC reaction) cycle proceeds as follows: mutational changes in receptor traits (Eq.~\ref{evo}) alter a cell's efficiency of acquiring antigen (Eqs.~\ref{EOM} and \ref{eta}), which in turn updates the selective advantage of competing clones (Eq.~\ref{fitness}), yielding a new generation of cells with modified binding characteristics.

\section{Results}
%Cells can sense their mechanical environment [] and use forces to guide their motion [] and fate decisions []. Immune cells appear to do so too~\cite{Wan:2013, Shaheen:2017, Hoogeboom:2018, Wang:2018}, but less is known about whether and when force usage confers a functional benefit. 
The observation that maximum antibody affinities evolved in organisms are considerably lower than those achieved by directed evolution in the laboratory indicates the presence of \textit{in vivo} factors that limit the efficacy of selection. We propose that tug-of-war antigen extraction holds the key to, at least in part, explaining this contrast by clarifying the impact of nonequilibrium recognition. In what follows, we identify the origin and determinants of affinity ceiling, evaluate to what extent active force usage influences selection, and propose ways in which cells may exploit physical constraints (e.g. antigen tethering) to their advantage.

\subsection{A. Antigen-tether strength sets affinity ceiling}
In a tug-of-war configuration (illustrated in Fig.~\ref{model}B), a B cell pulls on BCRs bound to antigens that are in turn tethered to the APC. Force propagates through the BCR-Ag-APC complexes, deforming coupled binding interfaces and modulating their lifetime distributions. The chance of antigen extraction, $\eta$, reflects the relative strength of the tugging and tethering bonds, measured by the ratio of expected lifetimes under force, $\tau_b$ and $\tau_a$, respectively. 
Defining relative tether strength $s\equiv\tau_a/\tau_{b0}$, where $\tau_{b0}$ denotes a founder B cell's BCR-Ag bond lifetime under force, the extraction probability becomes $\eta=\left[1+s/(\tau_b/\tau_{b0})\right]^{-1}$.
%; for founder B cells, $\eta=1/(1+s)$. Thus, strong tethers (large $s$) suppress antigen extraction, as seen in live B cells~\cite{Spillane:2017}.

As one would expect, strong tethers (large $s$) suppress antigen extraction. This has indeed been observed in live B cells~\cite{spillane2017}. More importantly, this expression implies a limiting factor on affinity maturation: 
%through rounds of mutation and selection, $\tau_b$ increases and $\eta$ rises. 
As $\tau_b$ increases well past $\tau_a$, $\eta$ tends toward saturation at $\eta_\mathrm{th}$ -- an extraction threshold above which selectable differences among cells become too little to drive further improvement in binding quality. Thus, affinity maturation hits a ceiling. 

\begin{figure}
\includegraphics[width=1\linewidth]{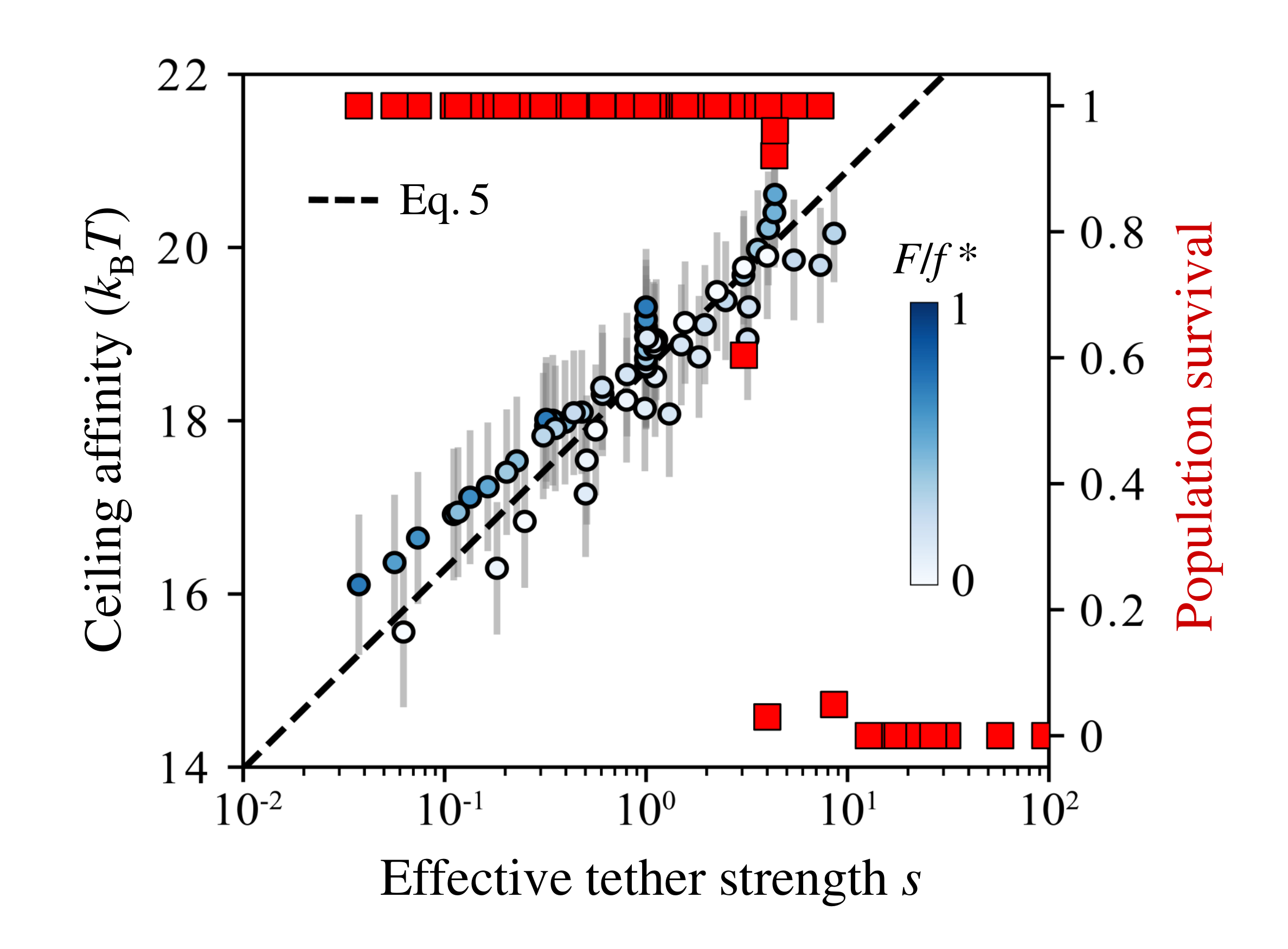}
\caption{Effective tether strength sets evolvable antibody affinity. 
Ceiling affinity -- evolved activation energy or barrier height $\Delta G_b^\ddagger$ -- largely follows a logarithmic dependence on relative tether strength $s\equiv \tau_a/\tau_{b0}$, as estimated by Eq.~5 (dashed line). Symbols are obtained from simulations over a wide range of tether strengths realized by varying the tether rupture length ($x_a^\ddagger: 0.5$--$4$nm) and force magnitude ($F: 0$--$30$pN); each symbol results from 100 runs for a given pair of $x_a^\ddagger$ and $F$ values. Modest deviation of symbols from the straight line arises due to relatively strong forces (darker blue circles) scaled by the critical force $f^* = \min\{3\Delta G_a^\ddagger/(2x_a^\ddagger), 3\Delta G_b^\ddagger/ (2x_b^\ddagger)\}$ at which the barrier to rupture vanishes. While the ceiling affinity rises with tether strength (circles), the fraction of surviving populations (red squares) vanishes quickly above tether strengths a few fold stronger than the founder BCR-Ag bond. Here, binding affinity $\Delta G_b^\ddagger$ evolves, while the bond length $x_b^\ddagger$ remains fixed. 
$\Delta G_a^\ddagger=\Delta G_{b0}^\ddagger=14k_\mathrm{B}T$, $x_b^\ddagger=2$nm. A linear-cubic potential is used here and after.
}
\label{ceiling}
\end{figure}

To make this intuition quantitative, we estimate the ceiling affinity $\Delta G_b^\ddagger$ by inverting the relation $\eta(\tau_b(\Delta G_b^\ddagger))=\eta_\mathrm{th}$ and compare to the output of simulated GC reaction. When activation barriers are high and forces modest, ceiling affinity follows a logarithmic dependence on tether strength
\begin{equation}\label{log_dep}
\Delta G_b^{\ddagger} \approx \Delta G_{b0}^\ddagger + k_{\rm B}T\left[\ln s + \ln \left(\frac{\eta_\mathrm{th}}{1-\eta_\mathrm{th}}\right)\right],
\end{equation}
where $\Delta G_{b0}^\ddagger$ is the founder affinity. As shown in Fig.~\ref{ceiling}, evolved affinities of simulated population ensembles (circles) match the prediction (dashed line) over a wide range of tether strengths, realized by systematically varying force magnitude ($F$: 0--30pN) and tether bond length ($x_a^\ddagger$: 0.5--4nm). 
%In these simulations, we focus on the evolution of binding energy $\Delta G_b^\ddagger$ while keeping the bond length $x_b^\ddagger$ fixed and identical among cells. Each circle in Fig.~\ref{ceiling} represents the evolved BCR affinity (final barrier height $\Delta G_b^\ddagger$) averaged over 100 populations for a given pair of $x_a^\ddagger$ and $F$.
Mild deviation arises under relatively strong forces (being a fraction of the critical force $f^*=\min\{3\Delta G_b^\ddagger/(2x_b^\ddagger), 3\Delta G_a^\ddagger/(2x_a^\ddagger)\}$ at which the barrier to rupture vanishes); in this regime, considerable landscape deformation can cause a nonlinear reduction in the log escape times, an effect neglected in our estimate.

Notably, force can modulate antigen tether strength and alter the ceiling.   
In fact, force causes differential influence on coupled bonds, depending on their relative stiffness and affinity. To the leading order in $F$, $s\sim\exp[F(x_b^\ddagger-x_a^\ddagger)/k_\mathrm{B}T]$, which indicates that tugging forces may enhance a stiff tether ($x_a^\ddagger<x_b^\ddagger$) but weaken a soft tether ($x_a^\ddagger>x_b^\ddagger$), yielding an elevated and a lowered affinity ceiling, respectively (Fig.~S1A). The intuition is that the same force, transmitted along a chain, would more strongly impact the softer bonds. As a consequence, pulling harder against stiff tethers/APCs will effectively strengthen the tether, suppress extraction, and therefore raise the affinity ceiling. Yet, this also raises the risk of population collapse (Fig.~S1B); strong tethers could lead to a deep population bottleneck and vanishing population survival (Fig.~\ref{ceiling}, red squares). These predictions are consistent with the observation that GC B cells undergoing affinity maturation exert strong forces against stiff APCs and extract fewer antigens compared to naive cells that use weak forces; force usage was found to enhance discrimination stringency at the cost of absolute extraction~\cite{nowosad2016}. As shown in Fig.~\ref{ceiling}, this tradeoff between affinity and survival yields an optimal tether strength a few fold stronger than the founder BCR-Ag bond.
%for parameter values informed by experiment.

We note that a precise prediction of the ceiling affinity requires full knowledge of tether properties, force magnitude, and founder affinity (Fig.~S1C). Nonetheless, a good catch of the overall trend by Eq.~\ref{log_dep} confirms that tether strength -- a composite parameter summarizing features of molecular interactions and mechanical environments as well as the impact of force -- indeed sets a bound to evolvable receptor affinities and at the same time reveals potential means to altering it.

\begin{figure*}
\includegraphics[width=1\textwidth]{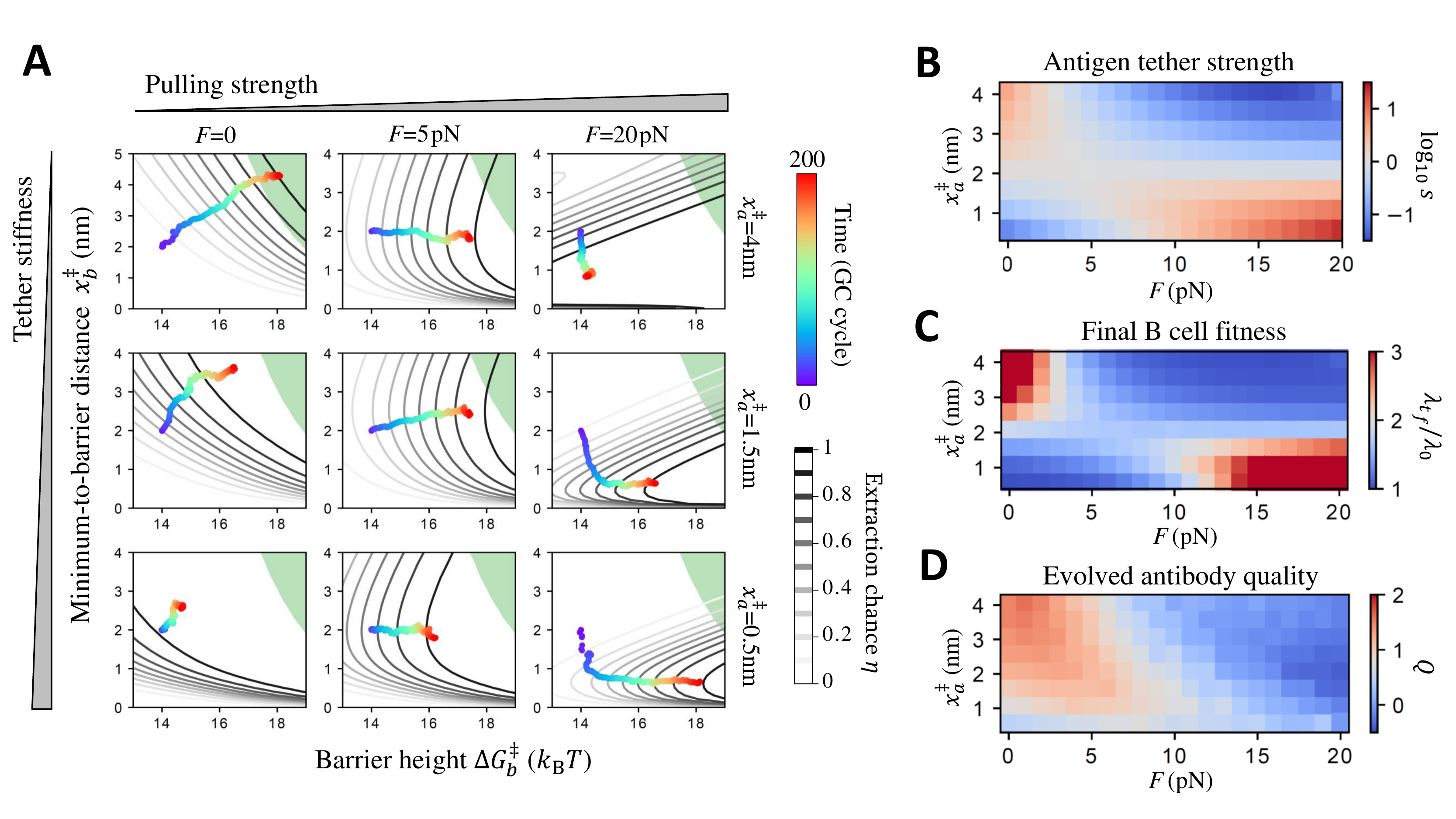}
\caption{
Active force usage steers B cell evolution and causes discrepancy between B cell fitness and antibody quality.
%conditions of training and testing for antigen recognition. 
(A) Example evolutionary trajectories of population-mean binding affinity $\Delta G_b^\ddagger$ and bond length $x_b^\ddagger$, guided by fitness landscape (contour map of extraction probability $\eta$) and subject to stochasticity in mutation and reproduction. The green region indicates high intrinsic binding quality ($Q\geq\log_{10}50$). Stronger pulling (left to right: $F=$ 0, 5pN, 20pN) selects for stiffer BCR-Ag bonds (smaller $x_b^\ddagger$), whereas varying tether stiffness (top to bottom: $x_a^\ddagger=$ 4nm, 1.5nm, 0.5nm) tunes the range and steepness of fitness gradient. Evolved BCRs appear to match the stiffness of antigen tethers; see panels along the diagonal. 
(B) Effective tether strength as a function of force magnitude $F$ and tether bond length $x_a^\ddagger$, obtained from mean first passage time calculation. Stronger forces strengthen stiff tethers ($x_a^\ddagger<x_b^\ddagger$) but weaken soft tethers ($x_a^\ddagger>x_b^\ddagger$), yielding maximum tether strengths at low force/soft tether and high force/stiff tether (red regions). 
(C) Dependence of evolved B cell fitness (fold increase in reproduction rate) on pulling force and tether stiffness largely follows the trend of tether strength (panel B). 
(D) Evolved antibody quality increases toward weak force/soft tether (upper left corner) just as evolved fitness does (panel C). However, strong force/stiff tether (lower right corner) lead to high B cell fitness yet low antibody quality. 
In (C) and (D), each pixel represents an average over 20 simulations, with equal mutation rates for $\Delta G_b^\ddagger$ and $x_b^\ddagger$; initial conditions are $\Delta G_{b0}^\ddagger=\Delta G_a^\ddagger=14k_\mathrm{B}T$ and $x_{b0}^\ddagger=2$nm.
}
\label{mismatch}
\end{figure*}

%\subsection{Force usage and stiffness sensitivity direct B cell evolution}
\subsection{B. Mismatch between B cell fitness and antibody quality reveals alternative functional objectives}
%training-testing discrepancy
We now explore in what ways the ability of cells to generate force and sense stiffness might influence their course of evolution. 
Fig.~\ref{mismatch}A demonstrates typical simulated trajectories (color-coded for time) on a fitness landscape calculated from extraction probability (contour map) in the 2D trait space of BCR affinity $\Delta G_b^\ddagger$ and bond length $x_b^\ddagger$. Force magnitude governs the overall direction of evolution: under vanishing force (left column), populations evolve toward softer BCRs (larger $x_b^\ddagger$), because soft bonds have a longer intrinsic lifetime in the absence of force. In contrast, strong pulling (right column) drives evolution toward stiffer BCRs (smaller $x_b^\ddagger$), since they are more resistant to force-induced barrier reduction, thus being longer-lived to support antigen extraction.   
For a given pulling strength, tether stiffness ($\sim 1/x_a^\ddagger$) tunes the steepness of fitness gradient. Most significant gradients occur at the corners -- low force/soft tether (upper left) and high force/stiff tether (lower right) -- consistent with maximum tether strengths (red corners in Fig.~\ref{mismatch}B), which lead to most pronounced increases in B cell fitness (Fig.~\ref{mismatch}C red corners) but at the expense of frequent population collapse (Fig.~S1B).  

Interestingly, it appears that BCRs evolve to match the tether/APC stiffness (Fig.~\ref{mismatch}A): soft presenting receptors/surfaces select for soft BCRs without force (upper left corner), whereas stiff substrates favor stiff BCRs under pulling (lower right corner). We propose that this ``stiffness mimicking" behavior might result from tug-of-war extraction in combination with mechanical feedback: stiffer substrates preferentially select cells exerting stronger contractile forces (Fig.~S2A, red diagonal), which in turn favor the usage of stiffer BCRs for antigen acquisition (Fig.~S2B). This proposal can potentially be tested by altering APC stiffness, via cholesterol depletion or induced inflammation, and measuring changes in evolved BCR-Ag bond length using single-molecule pulling experiment.    
In addition, our result predicts that rigid BCRs/antibodies evolve under strong tugging forces while flexible ones prevail under weak pulling. Through affinity maturation, some classes of antibodies become more rigid as affinity increases~\cite{thorpe2007}, while others may exhibit a diverging trend of changes in flexibility~\cite{klein2013, jeliazkov2018}. It will be interesting to search for the predicted force-rigidity correlation among B cells/antibodies raised against antigens with different physical attributes (e.g. size, shape, charge).   

Intuitively, one would expect that learning is only effective if training and testing are performed under similar conditions. However, B cells are trained to recognize membrane-bound antigens via nonequilibrium extraction, while antibody quality is tested through equilibrium binding to free antigens. 
We thus expect that this discrepancy, by causing a distinction between the training objective (efficient antigen extraction) and the testing criterion (strong equilibrium binding), would result in a mismatch between B cell fitness and antibody quality.

To characterize this mismatch, we define binding quality $Q\equiv\log_{10}\tilde{\tau}_b/\tilde{\tau}_{b0}$ that measures the fold change in intrinsic (i.e. force-free) lifetime of the BCR-Ag bond as a result of evolution and contrast it with B cell fitness $\lambda$ (Eq.~\ref{fitness}). 
Comparing Figs.~\ref{mismatch}C and \ref{mismatch}D, we see that pulling against stiff tethers/APCs -- characteristic of evolving cells -- indeed very effectively enhances clonal fitness (Fig.~\ref{mismatch}C, lower right red blob), but the resulting antibody quality is low (Fig.~\ref{mismatch}D). Essentially, force alters the fitness landscape, ``misleading" the population to turn away from the region of optimal binding quality (green corners in Fig.~\ref{mismatch}A); see an example in the lower right panel of Fig.~\ref{mismatch}A. 
This result suggests a surprising possibility: force usage by evolving cells may not simply be optimizing receptor binding quality to the current target, because pulling may even reduce the relative fitness of intrinsically strong binders. Then, what might the immune system gain from the apparent loss in performance? In Section D, we propose an unexpected answer, which reveals a basic tradeoff and a possible balance.

\subsection{C. Dynamic selection pressure improves response quality}
As was shown above, tether strength under force limits evolvable affinity (section A), whereas distinct conditions for training and testing may retain low-affinity clones (section B). One might wonder, can immune cells alleviate these constraints? We examine two biologically plausible strategies for their capacity to improve the response quality: renewable tether and dynamic force. Both schemes result in time-varying selection pressure.  

\begin{figure*}
\includegraphics[width=1\textwidth]{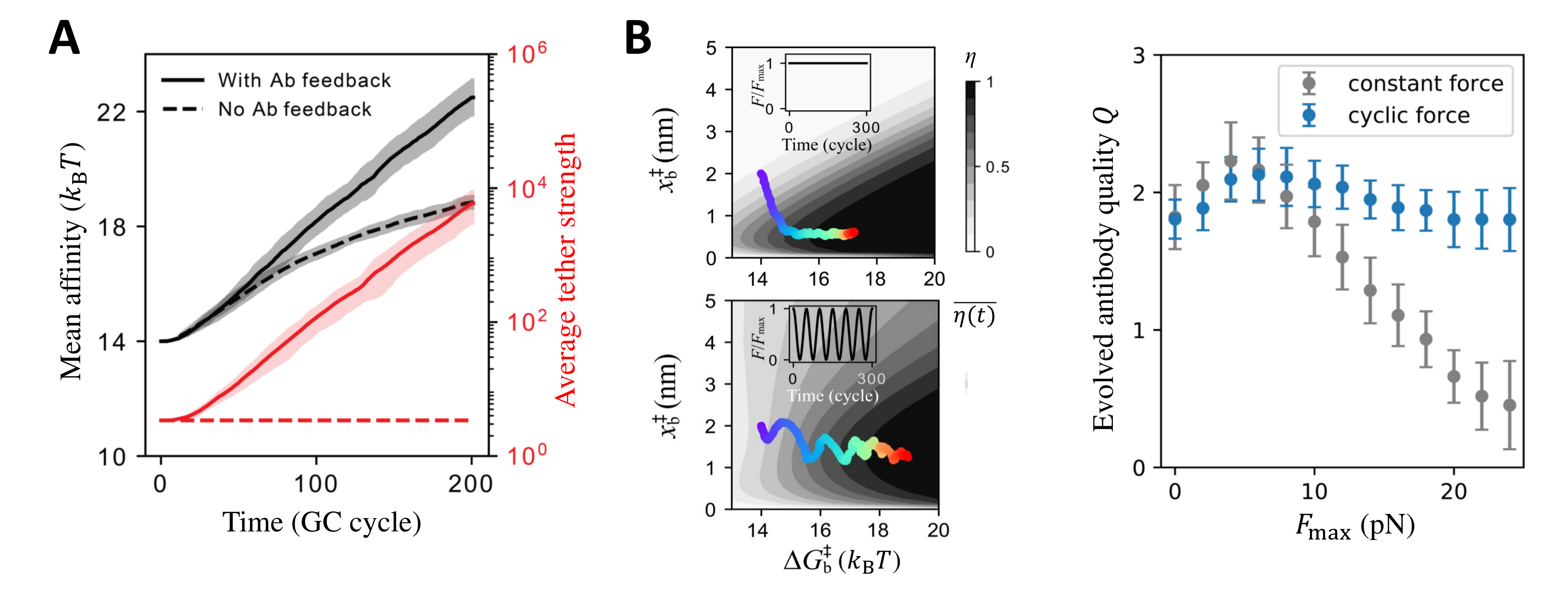}
\caption{Dynamic selection pressure sustains adaptation and improves evolved binding quality.
(A) Evolutionary dynamics with (solid lines) and without (dashed lines) antibody feedback. With fixed antigen tethers (red dashed line), the rate of increase in BCR affinity slows down over time (black dashed line). With renewing tethers sampled from high-affinity plasma cells (top-K rank in $\Delta G_b^\ddagger$), the overall tether strength (red solid line) and BCR affinity (black solid line) improve at similar steady rates, indicating sustained adaptation. The shade shows variation among 10 simulations. $\Delta G_a^\ddagger=14 k_\mathrm{B}T$, $x_a^\ddagger=1.5$nm, $x_b^\ddagger=2$nm; $F=20$pN; K=100.   
(B) Population-average evolution trajectories (left) and outcomes (right) under constant and oscillatory forces. Under constant forces, the fitness landscape (contour map of extraction probability) has a shallow affinity gradient and drives evolution toward stiff BCR-Ag bonds (upper left panel),leading to a rapid fall in evolved binding quality with increasing force magnitude (right panel, black symbols). Under not too slow oscillatory forces, populations evolve in an effective time-averaged fitness landscape (lower left panel) that has an attractor at high affinity and matching stiffness ($x_b^\ddagger\sim x_a^\ddagger$), resulting in high quality over a wide range of force magnitude (right panel, blue symbols). 
Constant force: $F=20$pN; oscillatory force: $F=F_{\rm max} (1+\cos(2\pi t/T_F))/2$, with $F_{\rm max}$=20pN, $T_F=50$; $t_f=300$. Error bars indicate variation among 20 simulations. $\Delta G_a^\ddagger=14k_\mathrm{B}T$, $x_a^\ddagger=1.5$nm; $\Delta G_{b0}^\ddagger=14k_\mathrm{B}T$, $x_{b0}^\ddagger=2$nm.
}
\label{dynamic}
\end{figure*}

Once BCR-Ag bonds become nearly as strong as the tethers, selection pressure vanishes and population affinity approaches saturation. Thus, key to lifting the affinity ceiling is the ability to strengthen the tether at a steady pace, neither too fast nor too slow, but best to match the rate at which BCR affinity improves. In this way, populations of cells adapt to the toughening environment at their best ability, neither being slowed by saturating extraction nor by severe population bottleneck. 
We test this idea \textit{in silico}: at the end of each GC cycle, we sample from the high-affinity members (with top ranks in $\Delta G_b^\ddagger$) of a cumulative plasma cell population to form a pool of feedback candidates. From this pool, a random subset is drawn to supply antibodies as antigen tethers for the next cycle. This setting of antibody feedback is motivated by \textit{in vivo} experiment showing that passively injected antibodies of high affinities can replace endogenous tethers to present antigen in mouse lymph nodes~\cite{zhang2016, toellner2018}. 

In Fig.~\ref{dynamic}A we compare the course of affinity maturation with and without antibody feedback. With fixed tethers (dashed lines), BCR affinity first rises then levels off. With renewing tethers (solid lines), however, contemporary clones receive a negative feedback from high-performing predecessors and achieve persistent adaptation. 
Importantly, overall tether strength (red solid) and average BCR affinity (black solid) indeed increase at a similar and steady rate, reflecting a restoring effect of antibody feedback: As effective tether strength becomes steady, extraction chance stabilizes, followed by a stable population size and selection strength, and hence a steady rate of adaptation. 
%This intuition and its interpretation are further discussed in a companion paper [].
Eventually, antigen tethers might become so strong that the weakest link in the chain shifts to the membrane tube pulled out by the tugging force, which ultimately limits evolvable affinities; in this case, further increases in BCR-Ag affinity make no more difference in extraction hence fitness. 

An apparent drawback of pulling against stiff APCs is that it inevitably drives selection of stiff BCR-Ag bonds that are short-lived without force (Fig.~\ref{dynamic}B top panel) -- the condition under which antibodies detect pathogens. As a possible remedy, cells may attempt to evolve and maintain $x_b^\ddagger$ to near $x_a^\ddagger$. This is because with matching stiffness, extraction is almost independent of force (to the leading order, $\tau_a/\tau_b\sim\exp[F(x_b^\ddagger-x_a^\ddagger)/k_\mathrm{B}T]$), directly resolving the training-testing discrepancy. 
One way to achieve this is to apply pulling forces in an oscillatory manner: $F(t)=F_\mathrm{max}(1+\cos(2\pi t/T_F))/2$; the oscillation period $T_F$ should be relatively short compared to the training duration ($t_f$) for ``dynamic localization" to be effective (see an example trajectory in Fig.~\ref{dynamic}B lower left panel). 
Such periodic modulation of pulling strength is plausible, through cyclic resetting of the cytoskeletal contractile machinery or via coupling to circadian rhythms of other cellular and organismal processes.
%This is plausible since the timescale of motor-induced contraction pulses ($\sim 10$ sec) is way shorter than a GC cycle ($\sim$ 4-6 hr).

In the fast-cycling limit, a population evolves in an effective fitness landscape that time averages those under varying force magnitudes. This effective landscape has an attractor at high affinity (large $\Delta G_b^\ddagger$) and matching stiffness ($x_b^\ddagger\approx x_a^\ddagger$), with a persistent gradient leading from low to high affinity (see contour map in Fig.~\ref{dynamic}B lower left panel). This is in stark contrast to the landscape under constant forces, for which affinity gradients are shallow and attractors biased toward extreme stiffness (Fig.~\ref{dynamic}B upper left panel, $F=F_\mathrm{max}$). Thus, as force increases in magnitude (Fig.~\ref{dynamic}B right), binding quality evolved under a constant force falls rapidly due to decreasing $x_b^\ddagger$ (black symbols). With cycling forces, however, response quality remains high regardless of force magnitude (blue symbols); clones forming too stiff ($x_b^\ddagger<x_a^\ddagger$) or too soft ($x_b^\ddagger>x_a^\ddagger$) BCR-Ag bonds are repeatedly removed during weak-force and strong-force periods, respectively, favoring the takeover by stiffness-matching clones ($x_b^\ddagger\sim x_a^\ddagger$) that remain fit under changing conditions.

\subsection{D. Heritable force heterogeneity diversifies binding phenotype: adaptive benefit of physical sensing}

Lastly, we ask whether immune recognition using forces can gain any advantage from the apparent loss in selected binding quality. 
Inspired by the observation of non-genetic variability among founder B cells~\cite{mitchell2018} -- resulting from intrinsic noise in molecular networks accumulated during stem cell differentiation -- we propose that heritable heterogeneity in force magnitude might be harnessed to generate diverse binding phenotypes. 
To test this hypothesis, we sample the force magnitude of founder cells from a distribution, assumed uniform for simplicity (with mean $F_\mathrm{ave}$ and width $\sigma_{F_0}$),
%$U[F_\mathrm{ave}-\sigma_{F_0}, F_\mathrm{ave}+\sigma_{F_0}]$, 
and analyze BCR traits resulting from evolution. Our predictions are robust to different choices of the distribution of founder force magnitude, as long as the values are limited to a finite range. One environmental cue capable of inducing variations in force magnitude is variability in APC stiffness, potentially caused by strain stiffening~\cite{hall2016} under B cell contractility and regulated by inflammatory signals~\cite{bufi2015}.
%; in fact, changes in tissue stiffness are known to feed back on force generation, e.g. by cancer cells [Mingming PNAS].

\begin{figure*}
\includegraphics[width=0.95\textwidth]{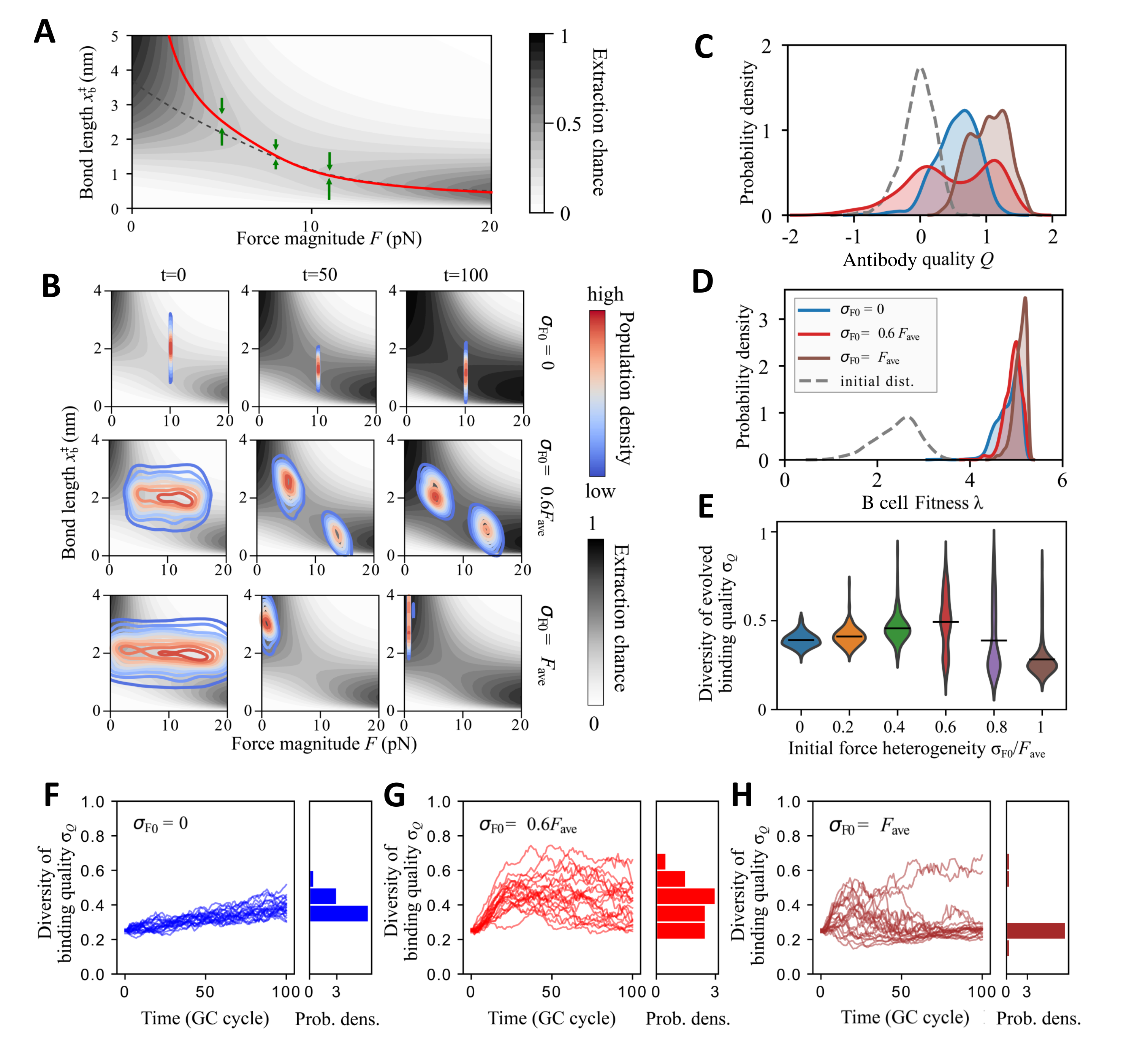}
\caption{An intermediate level of heritable force heterogeneity yields a broad range of binding quality and rate of diversity loss. 
(A) A saddle point in the fitness landscape (contour map showing extraction probability) is present at $F\simeq 9\mathrm{pN}$ and $x_b^\dagger=1.5\mathrm{nm}$. %Efficient antigen extraction is possible either with strong force and stiff bonds or with weak force and soft bonds; the latter has a higher absolute fitness. 
The ridge line (dashed black) traces along the direction of the positive principal curvature near the saddle point. The line of steady states (solid red), where the gradient in $x_b^\ddagger$ vanishes for a given $F$, deviates from the ridge line as force gets weaker, reflecting an asymmetry about the saddle point.
(B) Snapshots of evolving population density (color coded) in the $F$-$x_b^\ddagger$ space, starting from different levels of force heterogeneity. Extraction contours are evaluated with population-mean affinity at each time point. At an intermediate force heterogeneity (middle row, $\sigma_{F_0}=0.6F_{\mathrm ave}$), a population splits into two and remains bimodal for extended periods of time. Very small or very large heterogeneity (top and bottom rows) leads to a single crowd with similar force magnitudes. 
(C, D) Distributions of the evolved antibody quality $Q$ and B cell fitness $\lambda$ for zero (blue), intermediate (red), and maximum (brown) initial force heterogeneity; dashed lines indicate initial distributions. B cell lineages with diverse binding qualities (panel C) can have similar fitness (panel D) and coexist. (E) A violin plot shows that an intermediate force heterogeneity results in a broad range of diversity $\sigma_Q$ of evolved binding quality. Each violin was obtained from 20 simulations, with the horizontal bar indicating the ensemble average.
%some populations maintain bimodality for long while others quickly lose balance and become unimodal (panel G). At zero force heterogeneity (panel F), the diversity distribution remains narrow. Starting with very large heterogeneity of extraction force (panel H), the vast majority of populations exhibit a rapid loss of diversity in binding quality, with a tail of rare, long-lived bimodality. 
(F-H) Evolution of the diversity of binding quality, $\sigma_Q$, at zero (F), intermediate (G) and high (H) initial force heterogeneity.
The histogram on the side shows the final distribution of $\sigma_Q$.      
Parameters: $\Delta G_a^\ddagger = 14k_BT$, $x_a^\ddagger=1.5$nm, $F_{\rm ave}=10$pN, $t_f=100$. Mean and width of initial distributions: $\Delta G_{b0}^\ddagger=14k_\mathrm{B}T$, $\sigma_{0, G_b} = 0.2k_\mathrm{B}T$; $x_{b0}^\ddagger=2$nm, $\sigma_{0, x_b}=0.5$nm. Constant force, no antibody feedback. 
}
\label{saddle}
\end{figure*}

Fig.~\ref{saddle}A shows the fitness landscape (extraction probability as a proxy) as a function of force magnitude $F$ and bond length $x_b^\ddagger$ at population-mean affinity. Examination of the fitness contours identifies a saddle point at $F\simeq 9$pN and $x_b^\ddagger=1.5$nm. In its neighborhood, the fitness surface has a negative Gaussian curvature, with the ridge line (black dashed line) tracing the direction of the positive principal curvature. The presence of the saddle point is a direct consequence of the tug-of-war configuration and reflects the fact that efficient antigen extraction is possible for two different regimes: either large extraction force and stiff bonds or small force and soft bonds; the latter has a higher absolute fitness. 
%Note that the saddle point is quite asymmetric, with the ridge line bending up to the left. 

The emergence of a saddle point in the $F$-$x_b^\ddagger$ space is a generic feature of the tug-of-war extraction system. Extraction probability exhibits a logistic nonlinearity:
%with respect to binding affinity: 
$\eta=\left[1+\exp\left(\beta\Delta\Delta G+\ln(\tau_{s,b}/\tau_{s,a})\right)\right]^{-1}$, where the gap in activation barrier between the tugging and tethering bonds, $\Delta\Delta G=\Delta G_b^\ddagger(1-F/f_b)^{3/2}-\Delta G_a^\ddagger(1-F/f_a)^{3/2}$, and the ratio of time to leave the transition states, $\tau_{s,b}/\tau_{s,a}\propto \sqrt{(f_a-F)/(f_b-F)}$, have opposite dependence on bond length $x_b^\ddagger$ (via $f_b=3\Delta G_b^\ddagger/2x_b^\ddagger$) and force magnitude $F$, respectively (SI text); a saddle point occurs where the dependencies flip sign. Specifically, the $F$-dependence changes sign at $x_b^\ddagger\simeq x_a^\ddagger$. Meanwhile, when force is weak, time to escape the transition state dominates the $x_b^\ddagger$-dependence, giving rise to the fitness optimum at soft bonds. Under strong force, activation barrier dominates the dependence in favor of stiff bonds, resulting in the competing fitness peak (Fig.~S3).         

Evolution of the antigen extraction system takes a surprising turn because of this saddle point. Since evolution primarily operates on the bond length and binding affinity rather than the extraction force, we assume that force remains constant along lineages descending form the same founder cell, i.e. fully inheritable. Green arrows in Fig.~\ref{saddle}A show examples of the direction and magnitude of fitness gradient under this restriction: if the initial state is above (below) the red line, then the bond becomes more rigid (flexible). This line of steady states (with vanishing gradient in $x_b^\ddagger$) is close to the ridge line for strong forces but above it for weak forces.
%, where the ridge line has a larger slope and thus yields a greater vertical fitness gradient.      

We now follow the evolution of cell populations on this landscape according to stochastic agent-based simulations. We start all populations with a certain amount of initial diversity in bond length and binding affinity, but vary the initial diversity of the extraction force, $\sigma_{F_0}$. Fig.~\ref{saddle}B shows typical examples starting close to the saddle point. The fitness landscape changes over time due to the evolution of binding affinity, which does not significantly affect the saddle-point topology. Under zero initial force diversity (top row), mean bond length shifts toward $x_a^\ddagger$ while the fitness increases. With intermediate force diversity (middle row), population distribution splits into two groups with comparable fitness, one with higher extraction force and shorter bonds and the other with lower force and longer bonds. This bimodal distribution is clearly a consequence of the presence of the saddle point in the fitness landscape. One might expect that this population split will be enhanced by increasing the initial force heterogeneity. As we increase $\sigma_{F_0}$ by a considerable amount (from 6pN to 10pN), the split has completely disappeared (bottom row); there is a single population with a weak average force and a large mean bond length. The asymmetry of the saddle point is responsible for this: if the force distribution is sufficiently wide, cells with a force in the few pN range will rapidly evolve toward the fitness maximum, which then outcompete cells with large extraction force and small bond lengths. 

Fig.~\ref{saddle}C compares the distributions of intrinsic binding quality $Q$ at the end of evolution for zero (blue), intermediate (red) and large (brown) force heterogeneity (dashed line showing the initial distribution). Note that the double-peaked distribution for intermediate heterogeneity extends toward very low intrinsic quality (negative $Q$ indicating values lower than the initial mean). One would expect that the case of largest force heterogeneity produces B cells with highest fitness, because it most strongly expands the fittest subpopulation. Comparing the final fitness distributions for three cases (Fig.~\ref{saddle}D), the brown curve indeed shows a higher peak at a slightly larger fitness than those of the red and blue curves. However, the widths of the three distributions exhibit a significant overlap. It appears that, because of the saddle point, lineages with diverse binding qualities can have similar fitness and are able to coexist. This thus provides an alternative explanation for the persistence of low-quality clones, even under strong selection.    

Since B cell populations evolve in concurrent germinal centers (GCs), to what extent does the diversity of binding quality vary from one GC to another? Fig.~\ref{saddle}E compares the outcome of population ensembles evolved under varying initial force heterogeneity. Shown are probability densities of the diversity of final binding quality, $\sigma_Q$, in a violin plot. For zero initial force heterogeneity, the distribution is approximately Gaussian; $\sigma_Q$ is narrowly distributed among populations. As force heterogeneity is increased, the distribution acquires a tail of populations with much larger diversity. Then, for intermediate force heterogeneity, the distribution becomes bimodal, with a very wide range of diversities. Among 1000 realizations, about 60\% remain bimodal until the end, while nearly equal proportions of the rest become either fully low force/soft bonds or fully high force/stiff bonds, reflecting stochasticity in mutation and reproduction. When force heterogeneity is increased further, the distribution returns to, approximately Gaussian, though still with a tail towards large diversity. Figs.~S4-S9 present temporal characteristics of subpopulations with varying force magnitude in different regimes.  

It is worth pointing out that heritability of force magnitude is essential for bimodality and the non-monotonic dependence on initial force heterogeneity. If force is non-inheritable (re-sampled from the initial distribution at each GC cycle), such that cells along a lineage may pull at different strengths, the evolved diversity of binding quality will remain low and insensitive to initial force heterogeneity (SI text, Fig.~S10). Indeed, Mitchell et al.~\cite{mitchell2018} found that inheritable variability among founder cells contributes much more to the heterogeneity in B cell fate than intrinsic noise during proliferation does. 

Finally, we examine the evolution of the diversity distribution (Fig.~\ref{saddle}F-H, Fig.~S11). At zero force heterogeneity, the diversity distribution remains narrow until the end (histogram on the side). For intermediate force heterogeneity, the rate of diversity loss varies widely among populations (Fig.~S11C), resulting in a wide variety of final diversity of binding quality (Fig.~\ref{saddle}G); \textit{in vivo} studies of GC dynamics have indeed reported widely disparate rates of diversity loss from GC to GC~\cite{tas2016}. When starting with a very large diversity of extraction force, there is initially a steep rise in diversity, indicating population split, similar to the case of intermediate heterogeneity. But lineages with low force/soft bonds soon take over, leading to a rapid loss in diversity of binding quality. 

Therefore, the presence of the saddle point permits -- for intermediate force heterogeneity -- evolution of B cell lineages with similar fitness but a very broad range of binding quality and a wide variety of force diversity. The key is that, for long periods, the trajectories remain restricted to the saddle-point region and this prevents evolution toward a single dominant peak in the fitness landscape. This result suggests a physical means by which energy-consuming microscopic processes 
can diversify cellular phenotype without compromising reproductive fitness. In this sense, tug-of-war antigen extraction might have evolved to balance the tradeoffs between the potency of response to the current pathogen and the breadth of coverage against future escape mutants or related pathogens. Such tradeoffs may stem from resource constraints in support of GC reaction and immune memory formation, maintenance and renewal.

\section{Discussion}

%Biological systems learn by accumulating changes in internal degrees of freedom -- subject to metabolic and physical constraints -- to respond better in future circumstances. 
%Different from learning of intelligent machines, biological learning is subject to metabolic and physical constraints inherent to operation of life forms.    
The adaptive immune system offers a unique opportunity for observing \textit{in vivo} rapid evolution of molecular recognition: the process of affinity maturation iteratively alters the B cell repertoire and yields functional readouts on molecular, cellular and organismic scales.
While specificity and potency are desirable receptor traits to evolve, B cell selection is not simply favoring strong equilibrium binding to antigens. Rather, how many times a B cell divides upon activation reflects its ability to physically acquire antigen through active molecular processes. 
As an attempt to identify the limit and potential of immune adaptation, we develop a theoretical framework that maps binding affinity to clonal fitness via antigen extraction. This framework allows us to explore how active forces and physical constraints shape selection pressure, revealing alternative functional objectives.

We found that antigen extraction via a molecular tug of war results in both limiting and enabling effect of active force usage. First, it modulates antigen tether strength and hence limits evolvable antibody affinities. Our theory, applying Kramers expression to the evolution of BCR-Ag interaction, provides a quantitative dependence of affinity ceiling on force magnitude and intrinsic properties of binding energy landscapes, via a composite parameter -- the effective antigen tether strength (Fig.~2, Eq.~4). Intuitively, once the BCR-Ag bond becomes stronger than the tether due to evolution, selection pressure vanishes; thus, by effectively strengthening the tether, pulling forces can sustain adaptive evolution. However, too strong pulling increases the risk of population collapse during a deep bottleneck. Therefore, tug-of-war signal extraction causes \textit{in vivo} physical constraints on evolution. This helps rationalize why directed evolution \textit{in vitro} can produce much stronger binders than those realizable in living organisms; the laboratory selection platform decouples the target molecular interaction from its native cellular environment, lifting the constraints associated with tethering.

The enabling role of active force usage lies in how it influences the adaptive potential against future threats. 
Interestingly, our model suggests that molecular noise in force generation can yield plastic binding phenotype: an intermediate level of force heterogeneity among founder cells enables the generation and maintenance of a broad range of affinities among coexisting lineages. This behavior, rather than simply attributable to ineffective selection, may arise from a saddle point in the fitness landscape of B cell selection, which allows a wide variety of binding phenotypes (combinations of force magnitude and bond flexibility) to have similar fitness and hence coexist for extended time. Saddle point topology, combined with constrained heterogeneity of force magnitude, may explain multiple experiments, including retention of low-quality clones, coexistence of lineages with varying affinities to current antigen, and diverging rates of diversity loss among GC populations. We note that our proposal does not exclude other contributing mechanisms. For instance, in the case of complex antigens composed of multiple epitopes, synergistic interactions among coevolving B cell lineages -- e.g. through binding to allosterically coupled epitopes -- can promote persistence of low-affinity clones~\cite{yan2020}. Response to non-native antigen forms may also contribute ~\cite{kuraoka2016}. 
%Finally, our result implies that APC stiffness and tether properties could modulate the relative abundances of selected clones that use strong force/rigid bonds and weak force/soft bonds. 

A tug-of-war extraction system also allows cells to sense and adapt to mechanical cues in the physical environment, utilizing the tethering interaction  to perform comparative measurements and create dynamic feedback. For example, APCs can alter their stiffness in response to inflammatory signals from the innate immune system~\cite{bufi2015}, or at a faster speed, via mechanical feedback such as strain stiffening~\cite{hall2016} under B cell contractility. The model predicts a ``stiffness matching" behavior in which selected BCRs mimic the stiffness of the tethering complex, suggesting affinity discrimination as a functional objective, because most sensitive discrimination between similar BCR affinities is achieved when tugging and tethering bonds match in stiffness.  

Immune memory formation upon repeated exposure to pathogens or their antigens is a topic of lasting interest and longstanding debate. The phenomenon of immune imprinting~\cite{cobey2017}, by which the immune repertoire is strongly directed toward the primary infecting strain even after the virus has drifted antigenically, appears to indicate highly effective reactivation of memory B cells. A recent prime-boost study in mice, however, showed that secondary responses are strongly restricted from reengaging the large diversity of memory B cells generated by priming, but are instead dominated by very few clones~\cite{mesin2020}. Our model suggests that variability in internal dynamics of cells might supply a persistent diversity of binding phenotype that compensates for the the apparent loss in genetic diversity upon GC reseeding. We speculate that this phenotypic route of diversity generation and maintenance can be advantageous: It permits efficiency in limiting viral harm while circumventing the cost associated with responding de novo. Moreover, in synergy with restricted clonality, it may mitigate self-confinement of immune repertoire due to back-boosting of existing memory clonotypes -- the phenomenon of original antigen sin that limits the efficacy of vaccines~\cite{vatti2017}. It does so by providing temporary immune coverage as new memories evolve to form from naive ancestors. In fact, recent modeling work~\cite{wang2017} suggests that loss in clonal diversity upon boosting can favorably support the expansion and dominance of cross-reactive clones that target invariant viral features, under serial exposure to distinct but related antigen variants.

The emergence of phenotypic plasticity is expected to bear multiple observable signatures. First, the predicted correlation between force magnitude and bond flexibility (high force/stiff bond and low force/soft bond) can be sought among matured B cells exported from individual GCs. 
By collecting B cells from multiple GCs at the start and end of a response, one can test the non-monotonic dependence of evolved diversity in binding quality on initial force heterogeneity. Further, if time series data can be obtained from longitudinal tracking of diversity in force, slowest loss in diversity would be expected at intermediate levels of initial force heterogeneity.

%The proposed diversifying mechanism via a phenotypic route is likely most relevant to short-term immune coverage by nascent memory cells. For longer-term protection against recurrent or persistent pathogens, however, directed specialization may be required to further enhance the potency of response. Indeed, memory cells can reenter germinal centers upon activation and mature further [], potentially allowing for lasting protection. In this dynamic regime of longer timescales, environmental statistics and dynamics may play a central role in determining the optimal strategies of memory formation, storage and renewal.

We demonstrated an unexpected mismatch between the conditions under which training and testing of B cells for antigen recognition are conducted,
%. While antibodies function through equilibrium binding to free antigen, clonal selection is based on force-modulated extraction of membrane-tethered antigen. This discrepancy 
suggesting that active sensing by cells may not simply optimize receptor potency against current target. In addition, this discrepancy suggests an asymmetry between antigenicity and immunogenicity, i.e., distinction between B-cell activation and antibody recognition, which was predicted to play a large role in determining the course and fate of viral-immune coevolution~\cite{jiang2019}.
To construct alternative cost functions in optimization schemes, a systematic understanding of the physical basis of immune sensing and adaptation is needed to characterize trade-offs between evolvable traits, such as force-stiffness and affinity-flexibility relations provided by our model. 
%Interpretable, biophysically-grounded cost functions will, in turn, enable principled discovery of learning and memory strategies to counter diverse and evolving pathogens. 
%Practical implications include how vaccines should be presented in space and scheduled in time to balance short-term resource costs associated with affinity maturation and long-term response efficacy.  

One omitted characteristic in our theory is cellular organization of contact patterns. GC B cells collect antigens into clusters and extract them using forces. It will be interesting to study dynamics of cluster rupture via synaptic contact, by incorporating antigen extraction to the active membrane model we earlier developed~\cite{knezevic2018}, and to examine potential trade-offs between speed and accuracy of affinity discrimination. 
This augmented framework will also allow us to investigate design principles of sensing structures and control algorithms under different, potentially conflicting, functional needs.
Other interesting aspects for future study include the extent to which the relative frequency of mutations in antigen-engaging and framework regions, and the degree of force heritability, influence the breadth and effectiveness of responses, in both constant and fluctuating environments. We hope that our analysis of a minimal model of immune adaptation will motivate further work on conditions under which biological systems adapt  by exploiting physical influences on function and evolution. 

\section{Acknowledgement}
We thank Robijn Bruinsma for enlightening discussions. We are grateful for funding support from the Bhaumik Institute for Theoretical Physics at UCLA, the National Science Foundation (NSF) Grant MCB-2225947 and an NSF CAREER Award PHY-2146581. 

\bibliographystyle{unsrt}
\bibliography{main}

\end{document}